\documentclass[conference]{IEEEtran}
\IEEEoverridecommandlockouts

\usepackage{cite}
\usepackage{amsmath,amssymb,amsfonts}
\usepackage{algorithmic}
\usepackage{graphicx}
\usepackage{textcomp}
\usepackage{xcolor}
\usepackage{hyperref}
\usepackage[normalem]{ulem}
\usepackage{balance}
\usepackage{tabularx}
\usepackage{textcomp}

\def\BibTeX{{\rm B\kern-.05em{\sc i\kern-.025em b}\kern-.08em
    T\kern-.1667em\lower.7ex\hbox{E}\kern-.125emX}}
\begin{document}

\title{TFHE-Coder: Evaluating LLM-agentic Fully
Homomorphic Encryption Code Generation\\
}

\author{
Mayank Kumar\enskip Jiaqi Xue\enskip Mengxin Zheng\enskip  Qian Lou* \\
University of Central Florida \\}


\maketitle

\def\thefootnote{$*$}\footnotetext{ Corresponding Author. Email: qian.lou@ucf.edu.}

\begin{abstract}
Fully Homomorphic Encryption over the torus (TFHE) enables computation on encrypted data without decryption, making it a cornerstone of secure and confidential computing. Despite its potential in privacy-preserving machine learning, secure multi-party computation, private blockchain transactions, and secure medical diagnostics, its adoption remains limited due to cryptographic complexity and usability challenges. While various TFHE libraries and compilers exist, practical code generation remains a hurdle. We propose a compiler-integrated framework to evaluate LLM inference and agentic optimization for TFHE code generation, focusing on logic gates and ReLU activation. Our methodology assesses error rates, compilability, and structural similarity across open and closed-source LLMs. Results highlight significant limitations in off-the-shelf models, while agentic optimizations—such as retrieval-augmented generation (RAG) and few-shot prompting—reduce errors and enhance code fidelity. This work establishes the first benchmark for TFHE code generation, demonstrating how LLMs, when augmented with domain-specific feedback, can bridge the expertise gap in FHE code generation.
\end{abstract}


\section{Introduction}

Fully Homomorphic Encryption (FHE)~\cite{gentry2009fully, lou2019she, zhang2024heprune, lou2021safenet} allows computing over encrypted data, eliminating the need for decryption during processing~\cite{brakerski2014leveled, lou2021hemet, brakerski2012fully, zhang2023hebridge, fan2012somewhat, cheon2017homomorphic, chillotti2020tfhe, xue2022audit, zhang2025cipherprune,lou2019glyph, zheng2023primer}. It is, therefore, a promising cryptographic tool to ensure data privacy in the settings of secure computation, such as privacy-preserving machine learning~\cite{gilad2016cryptonets,lou2019autoq, santriaji2024dataseal}, secure multi-party computation~\cite{jin2023fedml}, private blockchain transactions~\cite{madathil2023prifhete}, and secure medical diagnostic~\cite{raisaro2018protecting}. There are various practical FHE schemes have been proposed. Among these, the TFHE scheme stands out. It is unique by offering efficient gate bootstrapping and functional bootstrapping, which allow for the computation of arbitrary functions while refreshing the noise. However, TFHE~\cite{deng2024trinity, zheng2024ofhe} poses significant challenges for developers, as it demands a high level of cryptographic expertise to configure encryption parameters (e.g., security levels, ciphertext dimensions) and optimize circuits for efficiency. These challenges make prototyping with TFHE cumbersome and hinder its broader adoption. Addressing these usability issues could substantially lower the entry barrier, enabling more researchers and developers to harness TFHE to prototype their ideas easily.


Recent advancements in Large Language Models (LLMs)~\cite{jiang2024survey,xue2024trojllm} have showcased their remarkable capacity to comprehend human language. Regarding coding, LLMs can assist developers by suggesting code snippets and even offering solutions to common programming challenges~\cite{mastropaolo2023robustness, nijkamp2022codegen}. To this end, leveraging LLMs' capabilities to assist developers implement TFHE application is a promising avenue for addressing the complexities associated with TFHE implementations. It would be valuable to evaluate if LLM could help developers with the TFHE coding, such as encryption parameters configuration and correct API calling. To this end, the expertise barrier could be significantly lowered, making TFHE more accessible for developer with few related expertise.


\begin{figure}[t]
    \centering
    \includegraphics[width=0.35\textwidth]{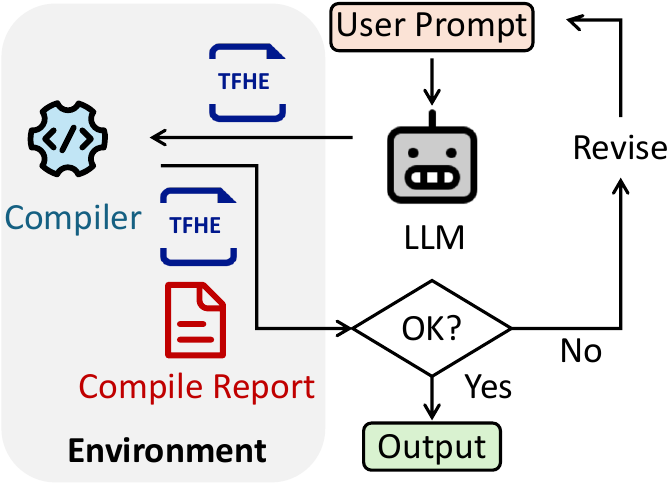} 
    \caption{Overview of the \textit{compiler-in-the-loop} evaluator. The LLM generates TFHE code based on a user prompt, which is then compiled. If compilation fails, the model receives a compile report and revises its output iteratively until a compilable solution is produced.}
    \label{fig:baseline}
\end{figure}

\begin{figure}[t]
    \centering
    \includegraphics[width=0.35\textwidth]{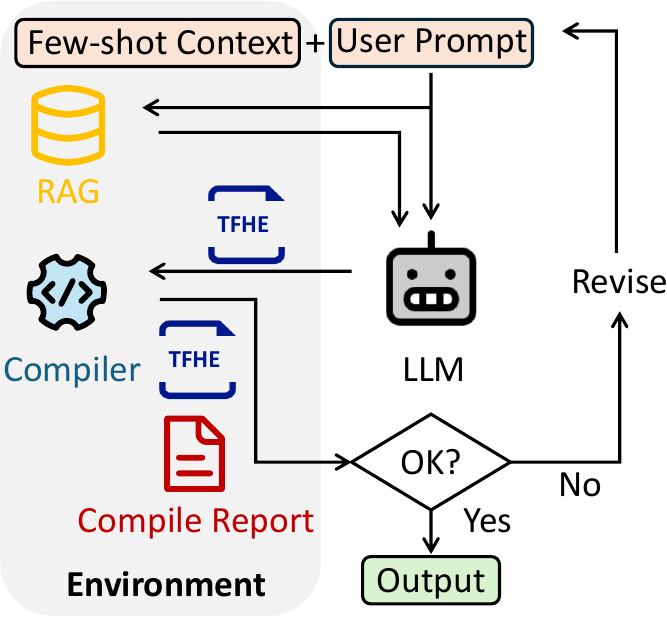} 
    \caption{\textit{Agentic-optimized} evaluator loop incorporating RAG and few-shot prompting. The LLM receives both retrieval-augmented documentation and few-shot examples alongside the user prompt, refining its output iteratively based on compile reports until a compilable solution is achieved.}
    \label{fig:improvement}
\end{figure}

Although LLMs excel in general programming tasks, generating code for domain-specific libraries and languages remains challenging~\cite{joel2024survey}. Despite TFHE’s significance in national security, secure computation, and privacy-preserving machine learning, no studies have evaluated LLMs for TFHE code generation. Given its programming complexity, even for cryptographic experts, assessing whether LLMs can assist in TFHE coding is valuable. A key concern is whether LLMs, given natural language task descriptions, struggle with correct API usage and tend to produce verbose or unfocused outputs. Additionally, the lack of standardized metrics for evaluating TFHE code correctness and functionality necessitates reliance on manual inspection. These challenges underscore the need for robust evaluation frameworks and stricter control over LLM-generated code to meet TFHE’s stringent security and development standards.


To systematically evaluate LLM performance in TFHE code generation, we propose Fig. \ref{fig:baseline}, the \textit{compiler-in-the-loop} workflow and Fig. \ref{fig:improvement}, the \textit{Agentic-optimized} workflow. The \textit{compiler-in-the-loop} workflow iteratively refines LLM-generated TFHE code by compiling the output and using compiler diagnostics for feedback. If the code compiles successfully, it is accepted; otherwise, the LLM revises its response based on the compile report, repeating the process until a valid output is achieved or the iteration limit is reached. The \textit{Agentic-optimized} workflow enhances the compiler-in-the-loop approach by integrating retrieval-augmented generation (RAG)~\cite{lewis2020retrieval, xue2024badrag,gao2023retrieval} and few-shot prompting~\cite{reynolds2021prompt, zhengtrojfsl, gu2021ppt}. The LLM leverages retrieved TFHE documentation and validated examples to refine its outputs iteratively, guided by compiler feedback, until a compilable implementation is achieved.

Our experiments focus on foundational tasks, specifically, functionally complete logic gates (NOT, AND, OR) and ReLU activation, which are selected for their relevance to cryptographic circuits and privacy-sensitive applications like machine learning, to assess code quality and identify common failure patterns. The evaluator loop simulates real-world debugging workflows by compiling the LLM-generated TFHE code and leveraging compiler diagnostics to detect syntax errors, while also keeping track of TFHE-specific issues. These include failures such as invoking irrelevant or non-existent TFHE library APIs, often resulting from limited TFHE training data or hallucination. Such errors present unique challenges when using LLMs for TFHE code generation.

We evaluate both closed-source (GPT-4o~\cite{hurst2024gpt}, Claude-3.5-Haiku~\cite{anthropic2024claude3addendum}) and open-source (CodeLlama-7B~\cite{roziere2023code}, Qwen2.5-Coder-7B~\cite{hui2024qwen2}, Deepseek-Coder-6.7B~\cite{guo2024deepseek}) mainstream LLMs on several dimensions: compilability, functionality, and closeness to ground truth. Specifically, we adopt \textit{Pass@k Comp.} and \textit{Pass@k Func.} to measure the proportion of generated code that compiles and executes correctly, respectively. Secondly, we use \textit{CrystalBLEU} to assess how closely a model's output resembles a correct reference implementation. This metric helps evaluate the potential usefulness of a generated solution, even when it is not immediately executable, as higher similarity often indicates that less effort is required to correct and functionalize the output.



Our evaluation reveals key insights into TFHE code generation. In the compiler-in-the-loop workflow, GPT-4o outperforms open-source models, while weaker models struggle with formatting errors, API misuse, and repetitive mistakes, especially for complex tasks like ReLU. In the agentic-optimized workflow, RAG alone offers limited gains, whereas few-shot prompting significantly improves correctness. The best results emerge from combining RAG and few-shot prompting, reducing errors and enhancing functional accuracy. However, arithmetic-heavy tasks like ReLU remain challenging, highlighting the need for further refinement in guiding LLMs for encrypted computation.


In summary, our contributions are two-fold: 
\begin{enumerate}
    \item \textbf{\textit{Compiler-in-the-loop} Evaluation:} We benchmark off-the-shelf LLMs on TFHE code generation for core tasks, revealing critical limitations in error rates, compilability, and structural fidelity.
    \item \textbf{\textit{Agentic-optimized} Evaluation:} We demonstrate how combining few-shot prompting~\cite{brown2020language} and RAG~\cite{lewis2020retrieval} reduces errors and improves code similarity through compiler-integrated feedback cycles.
\end{enumerate}


\section{Background}

The intersection of large language models (LLMs) and fully homomorphic encryption (FHE) presents a unique opportunity to democratize secure computation. While LLMs have shown remarkable prowess in code generation for mainstream languages, their application to specialized cryptographic libraries like TFHE remains unexplored. This section examines the potential of LLMs to generate TFHE code, leveraging their understanding of logical operations, and explores the unique characteristics of TFHE that make it both challenging and promising for automated code generation.

\subsection{LLM for Code Generation}\label{AA}
Code generation is a key application of large language models (LLMs), with models such as CodeGen~\cite{nijkamp2022codegen}, CodeX~\cite{chen2021evaluating}, and CodeT5~\cite{wang2021codet5} excelling in widely used languages like C, C++, Python, and Java due to the availability of extensive training corpora. However, generating code for specialized libraries like TFHE, implemented in C, presents challenges due to its cryptographic complexity and niche API. Recent studies on LLMs for High-Level Synthesis (HLS) and Register Transfer Level (RTL) design~\cite{thakur2023benchmarking, liao2024llms, xiong2024hlspilot} demonstrate that LLMs can effectively model logical operations such as AND and OR gates. Given that TFHE operations also rely on gate-level computations, it is reasonable to hypothesize that LLMs, with appropriate improvement techniques, could generate functional TFHE code by leveraging their learned logical reasoning capabilities.

\subsection{Fully Homomorphic Encryption over the Torus}

Fully Homomorphic Encryption over the Torus (TFHE)~\cite{jiang2022matcha} operates on boolean circuits using logical gates (NOT, AND, OR) with explicit noise management through bootstrapping after each operation. While TFHE requires adherence to strict security parameters, it offers several advantages over schemes like BGV~\cite{brakerski2014leveled, yudha2024boostcom} and CKKS~\cite{cheon2017homomorphic} in terms of practical implementation. TFHE's boolean circuit approach aligns more closely with traditional programming paradigms, making it easier for developers to conceptualize and implement encrypted computations. Its efficient gate-by-gate bootstrapping is faster and more straightforward to implement than the complex relinearization and modulus switching procedures required in BGV/CKKS. TFHE's deterministic noise management simplifies handling in code implementation, as noise is reset after each gate operation. Additionally, TFHE's structure allows for efficient hardware acceleration, potentially simplifying high-performance implementations. However, programming TFHE still presents challenges, including the need to carefully manage bootstrapping operations and adhere to specific security parameters to maintain the scheme's integrity.

\section{Our Method}
\subsection{Problem Definition}

We propose an iterative compiler-integrated framework \textit{Compiler-in-the-loop} Evaluation, for generating compilable TFHE code from natural language specifications and reference implementations. As shown in Fig. \ref{fig:baseline}, the system tasks the LLM with synthesizing cryptographically compliant TFHE operations (e.g., bootstrapped NOT/AND/OR gates) while adhering to strict security parameters. Operating in an agentic fashion, the LLM interacts dynamically with its environment, including the compiler (using the tfhe library flags) and TFHE documentation, to refine its outputs. Through successive refinement cycles, the framework leverages compiler diagnostics to resolve syntax errors and API misuse, progressively improving code compilability. To improve the code generation capability, we propose the \textit{Agentic-optimized} evaluation in Fig. \ref{fig:improvement}, a hybrid approach combining retrieval-augmented generation (RAG) with TFHE documentation and few-shot prompting with validated circuit templates ensures alignment between generated code and cryptographic constraints. This methodology bridges the gap between developer intent and FHE implementation rigor, enabling accessible yet correct privacy-preserving computation.

    
    Our experiments are designed to answer the following research questions (RQs):
        \begin{itemize}
            \item
            \textbf{RQ1:} How well do mainstream LLMs generate syntactically valid and functionally correct TFHE code from a natural language prompt and reference implementation? 
            \item
            \textbf{RQ2:} What are the common reasons that LLMs fail to generate correct TFHE code? 
            \item  \textbf{RQ3:}  How does providing additional TFHE-specific context—such as retrieval-augmented generation (RAG) and few-shot prompting—impact code generation accuracy and error rates?
        \end{itemize}

\subsection{Method Design}

The task space comprises functionally complete logical operations (NOT, AND, OR) and the ReLU activation function. The gate set was selected for its ability to represent arbitrary boolean circuits, a foundational requirement for TFHE-based cryptographic protocols. ReLU, a ubiquitous activation function in machine learning, serves as a practical test case for translating plaintext operations into TFHE-encrypted equivalents. These tasks collectively evaluate LLMs’ ability to synthesize both low-level cryptographic primitives and high-level machine learning components using TFHE’s gate-level programming paradigm.

We select the mainstream LLMs commonly used in recent code-related studies, including both open-source and closed-source LLMs. For open-source LLMs, we choose Code Llama~\cite{roziere2023code} (CodeLlama-7B-Instruct), Qwen2.5-Coder~\cite{hui2024qwen2} (Qwen2.5-Coder-7B-Instruct) and DeepSeek Coder~\cite{guo2024deepseek} (deepseek-coder-6.7b-instruct). For closed-source LLMs, we select the widely used GPT-4o~\cite{hurst2024gpt} and Claude-3.5-Haiku~\cite{anthropic2024claude3addendum}. For all studied LLMs, we set the temperature to 0.9 and top-p to 0.85. Note that, to mitigate issues stemming from the randomness of model generation, the experimental results presented in this paper are obtained by conducting five repeated experiments and averaging the results. 

\textbf{Experimental Setup.}
All experiments were conducted on a desktop PC equipped with 12th Gen Intel(R) Core(TM) i5-12600K, 64 GB RAM, an NVIDIA GeForce RTX 3090 GPU with 24 GB VRAM, and running Ubuntu 24.04 with Python 3.8 and CUDA 12.2. The embedding model used for RAG was \texttt{jinaai/jina-embeddings-v2-base-code} \cite{embeddingModel}.

Our framework employs two distinct metric categories to evaluate LLM-generated TFHE code: (1) Generation-phase metrics and (2) Evaluation-phase metrics. 

\textbf{(1) Generation-phase metrics} track mistakes the LLM makes during code generation, regardless of whether the code compiles or functions correctly. These include the \textbf{Wrong Format Error}, where the model disregards user instructions for structured output formatting, preventing automated code retrieval, and the \textbf{Repetition Error}, where the model repeatedly generates the same incorrect response despite iterative feedback, indicating a failure to refine its output. These metrics reflect the model's adherence to user instructions and its responsiveness to feedback in an iterative generation process. 

\textbf{(2) Evaluation-phase metrics} assess the correctness and usability of the generated code after it has been produced. \textit{CrystalBLEU} quantifies the structural similarity between the generated and reference implementations by comparing API usage. Unlike direct compilability measures, CrystalBLEU provides an execution-independent evaluation of code fidelity, mitigating the need for computationally intensive TFHE execution. \textit{Pass@k (comp)} measures the fraction of generated solutions that successfully compile, while \textit{Pass@k (func)} evaluates the fraction of compiled solutions that produce correct functional outputs across test cases:

\begin{equation}
    \text{pass@k} = 1 - \frac{\binom{n_t - n}{k}}{\binom{n_t}{k}}
\end{equation}
where \( n_t \) represents the total number of generated solutions, and \( n \) denotes the number of solutions that pass the evaluation criteria. 

For \textbf{compilability} (\textit{Pass@k (comp)}), a solution is considered "passed" if it compiles successfully without errors. 

For \textbf{functional correctness} (\textit{Pass@k (func)}), a solution is considered "passed" if it produces the expected outputs across multiple unit tests. In our evaluation, $n_t=5$ and $k=1$.

Together, these metrics provide a comprehensive evaluation framework, capturing both syntactic and functional correctness while identifying systematic generation errors in TFHE code synthesis.

\subsection{LLM Methods}
\begin{itemize}
    \item \textbf{Baseline:}
    The baseline \textit{compiler-in-the-loop} workflow (Fig. \ref{fig:baseline}) iteratively refines TFHE code generation using an LLM and compiler feedback. A user prompt (consisting of task description and reference \texttt{C} code) is passed to the LLM, which generates TFHE code saved to disk. The code is compiled, and if successful, the process halts. Otherwise, a revision prompt is created from the compile error report and fed back to the LLM for correction. This loop continues for up to 10 iterations, tracking error metrics and calculating structural similarity upon termination.
    \item \textbf{RAG:}
    Building on the baseline workflow, in the \textit{Agentic-optimized} workflow (Fig. \ref{fig:improvement}), retrieval-augmented generation (RAG) is introduced  to provide the LLM with relevant TFHE-specific context. Before generating code, the system retrieves snippets from the official TFHE documentation~\cite{tfhe2024api} and appends them to the system prompt. This ensures that the LLM has access to accurate API definitions and usage examples, reducing errors such as incorrect function calls or hallucinated dependencies.
    \item \textbf{Fewshot Prompting:} Separately, few-shot prompting is incorporated into the baseline workflow by including a correct implementation of an OR gate using TFHE’s \texttt{bootsOR} function as part of the first user prompt. This example serves as a structural guide for the LLM, enabling it to generate other gate-level operations (e.g.,NOT,AND) with improved syntax and parameter usage. By providing a concrete example, few-shot prompting helps reduce ambiguity in the generated code.
    \item \textbf{Combining RAG + Fewshot:} The final workflow combines RAG and few-shot prompting to leverage their complementary strengths. RAG provides domain-specific knowledge from TFHE documentation, ensuring accurate API usage, while few-shot examples offer structural guidance for implementing gate-level operations. Together, these methods enhance the LLM’s ability to generate syntactically valid and semantically meaningful TFHE code, significantly reducing errors and improving alignment with reference implementations during iterative refinement cycles.
\end{itemize}


\section{Evaluation Results}

To study the effectiveness of LLM-generated TFHE code, we analyze trends across \textit{CrystalBLEU}, \textit{Pass@k (comp)}, and \textit{Pass@k (func)} to identify potential causes of failure. A high \textit{CrystalBLEU} score combined with successful compilation and functional correctness (high \textit{Pass@k (comp)} and \textit{Pass@k (func)}) indicates strong alignment with reference implementations. However, deviations in these metrics reveal distinct failure patterns. If \textit{CrystalBLEU} is high, but the code does not pass functional tests, the model likely \textit{misused an API} (e.g., applying \texttt{bootsOR} instead of \texttt{bootsAND}). When both \textit{CrystalBLEU} and \textit{Pass@k (comp)} are low, the model either \textit{failed to use the correct API} or \textit{introduced non-existent/hallucinated function calls}, resulting in code that does not compile. In cases where compilation succeeds despite a low \textit{CrystalBLEU} score, the model may have \textit{found an alternative solution} or \textit{engaged in copy-paste behavior from reference implementation}. If \textit{Pass@k (comp)} is low but \textit{CrystalBLEU} is moderate, the generated code is often structurally sound but contains \textit{minor syntactic errors or missing parameters}. Lastly, a scenario where the model produces compiled but functionally incorrect code suggests \textit{the use of correct APIs, but for irrelevant tasks}, such as misapplying encryption functions in unintended contexts. These trends provide a structured approach for diagnosing model weaknesses and guiding future refinements in TFHE code generation.

\begin{figure}[t]
    \centering
    \includegraphics[width=0.48\textwidth]{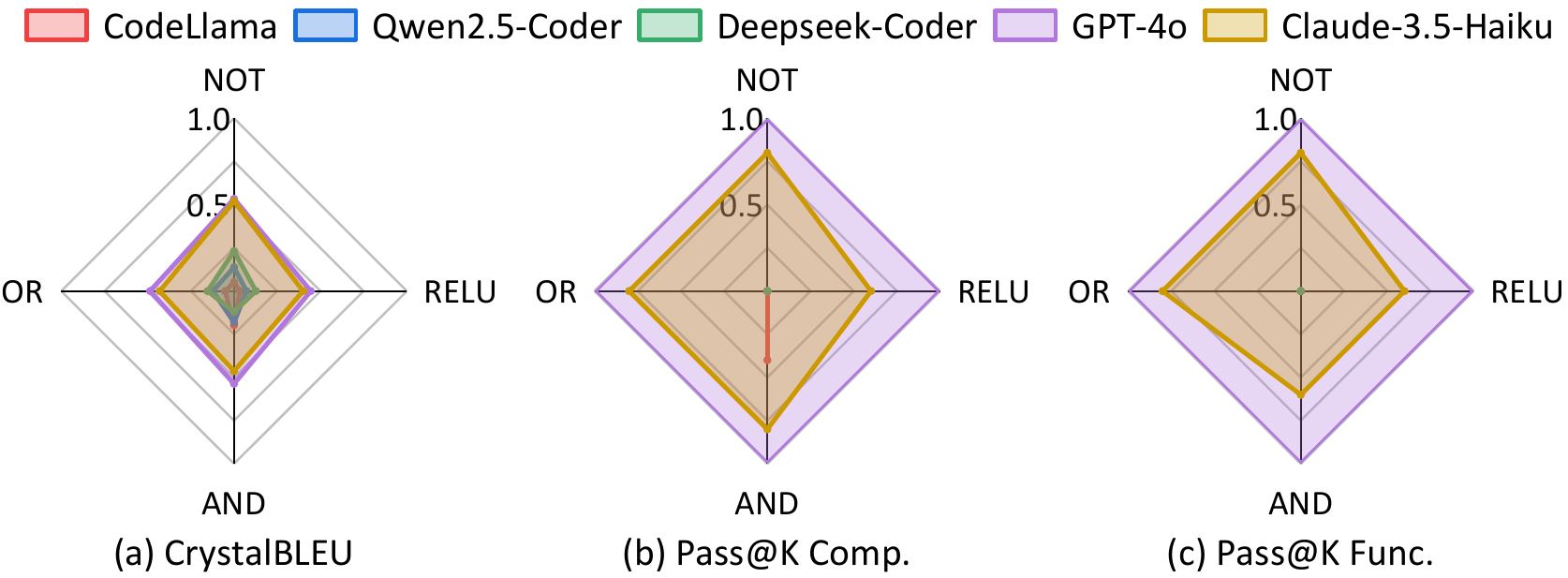} 
    \caption{Baseline performance comparison of all models across four tasks (NOT, AND, OR, ReLU) using: (a) CrystalBLEU, (b) Pass@k (comp), and (c) Pass@k (func). Higher values indicate better alignment with reference implementations, with GPT-4o consistently outperforming other models across all tasks.}
    \label{fig:baseline-plot}
\end{figure}

\noindent\textbf{RQ1:}
The results of baseline provide an overview of how mainstream LLMs perform in generating syntactically valid and functionally correct TFHE code from a prompt. As seen in Fig. \ref{fig:baseline-plot}, GPT-4o consistently outperforms others across all tasks, achieving the highest \textit{Pass@k (comp)} and \textit{Pass@k (func)} scores. Claude-3.5-Haiku follows closely, demonstrating moderate success in compilability, particularly for the NOT and OR gates, but struggles with full functional correctness.

Open-source models, including CodeLlama, Qwen2.5-Coder, and Deepseek-Coder, exhibit significantly lower performance. None of these models successfully generated functionally correct TFHE code, as indicated by their \textit{Pass@k (func)} scores of zero. This aligns with their low \textit{CrystalBLEU} scores, suggesting that their outputs diverge substantially from reference implementations.

Among tasks, ReLU appears to be the most challenging, with all models except GPT-4o failing to produce functionally correct implementations. This suggests that TFHE-based arithmetic operations introduce additional complexity beyond Boolean logic synthesis.

Overall, the results highlight a stark performance gap between close-source models like GPT-4o and Claude-3.5-Haiku versus open-source models in the baseline setting. This underscores the need for advanced techniques, such as RAG and few-shot prompting.

\noindent\textbf{RQ2:}
To understand why LLMs fail in TFHE code generation, we analyze the key error patterns observed in the baseline evaluation. The dominant failure cases can be categorized into three primary areas: (1) generation phase issues, including incorrect formatting and repetition errors, (2) evaluation phase failures, covering both syntactic and functional correctness issues.

\begin{figure}[th]
    \centering
    \includegraphics[width=0.48\textwidth]{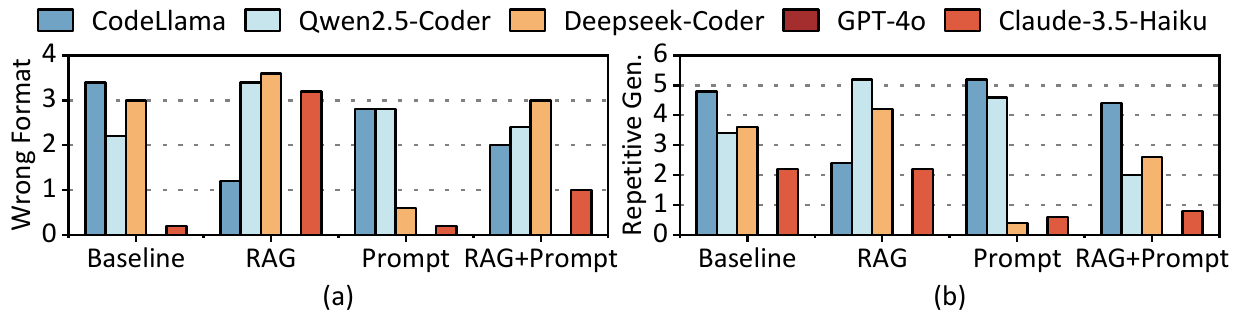} 
    \caption{Impact of Baseline, RAG, Few-shot Prompting, and their combination (RAG+Prompt) on (a) Wrong Format and (b) Repetition Error.}
    \label{fig:other-metrics-plot}
\end{figure}

\paragraph{Generation Phase Failures}
Considering Fig. \ref{fig:other-metrics-plot}, a significant proportion of failures stem from issues in how LLMs structure their responses. The \textit{Wrong Format} metric reveals that open-source models frequently fail to adhere to the requested response structure, making it difficult to extract usable TFHE code. Deepseek-Coder, for instance, exhibits the highest \textit{Wrong Format} rate across multiple tasks, indicating a systemic failure in following prompt constraints. Additionally, high \textit{Repetition Error} rates, suggest that models often regenerate the same incorrect solutions in multiple iterations, failing to incorporate compiler feedback for meaningful improvements.

\paragraph{Evaluation Phase Failures} 
When examining compilation and functional correctness, clear trends emerge. GPT-4o is the only model that achieves near-perfect \textit{Pass@k (comp)} and \textit{Pass@k (func)} scores across all tasks, whereas other models frequently fail both. Claude-3.5-Haiku achieves partial success in NOT and OR gates, but struggles with more complex tasks like ReLU. The most common reason for failed compilation among weaker models is the incorrect use of TFHE APIs, either due to missing parameters or hallucinated function calls. This is particularly evident in Qwen2.5-Coder and Deepseek-Coder, where low \textit{CrystalBLEU} scores (e.g., 0.0719 for ReLU in Qwen2.5-Coder) suggest that generated code diverges significantly from correct implementations.

\paragraph{Failure Trends Across Tasks.} 
Different tasks exhibit varying levels of difficulty. As shown in Fig~\ref{fig:baseline-plot}, logic gates such as NOT and OR generally yield higher \textit{CrystalBLEU} scores, indicating some structural alignment with reference implementations. However, despite this alignment, the \textit{Pass@k (comp)} and \textit{Pass@k (func)} scores remain low, implying that while models generate syntactically reasonable TFHE code, they often misuse APIs (e.g., applying \texttt{bootsOR} instead of \texttt{bootsAND}). In contrast, the ReLU task sees the lowest \textit{CrystalBLEU} scores across models, confirming that it is the most challenging due to its arithmetic complexity.

Overall The primary reasons for LLM failure in TFHE code generation can thus be attributed to structural formatting issues, repetitive incorrect generations, API misuse, hallucinated function calls, and an inability to generalize beyond template-based logic. Open-source models show particularly high rates of these failure patterns, whereas closed-source models, particularly GPT-4o, demonstrate superior robustness in addressing these challenges.

\begin{figure}[th]
    \centering
    \includegraphics[width=0.48\textwidth]{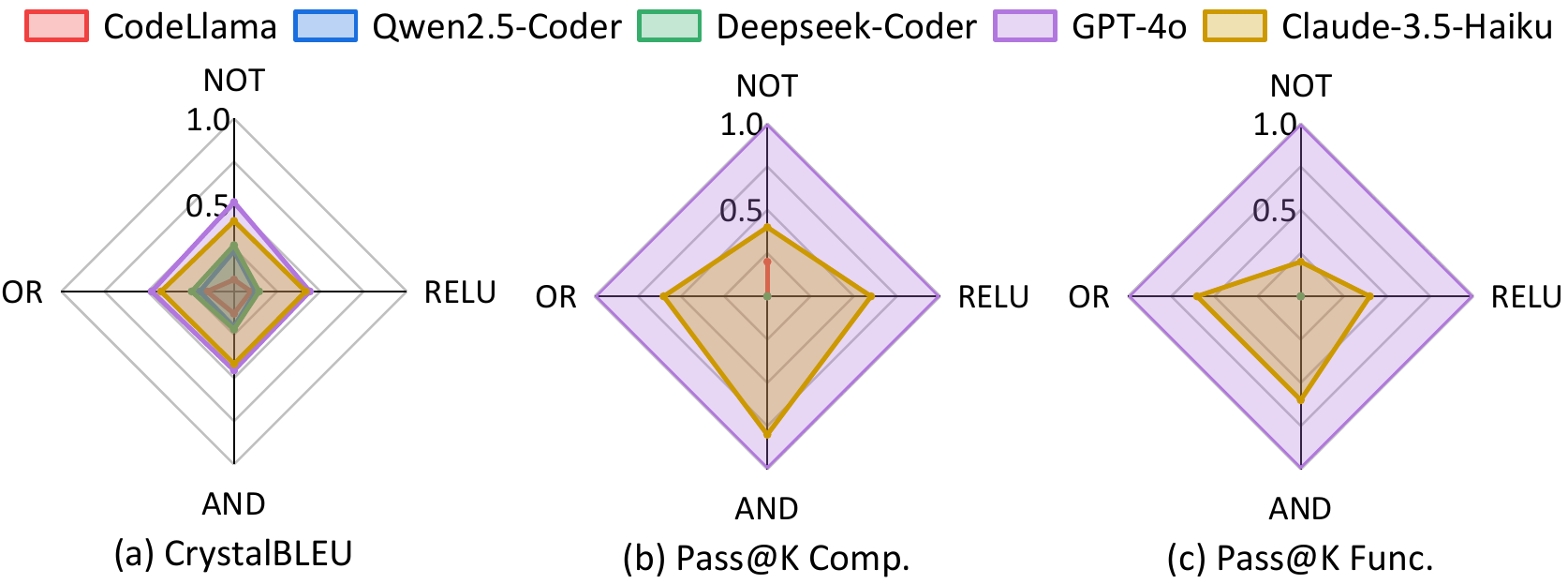} 
    \caption{Comparison of all models across four tasks using RAG technique, using: (a) CrystalBLEU, (b) Pass@k (comp), and (c) Pass@k (func). While minor improvements in CrystalBLEU are observed for some models, overall functional correctness remains low, with GPT-4o maintaining the highest performance across all tasks.}
    \label{fig:rag-plot}
\end{figure}

\begin{figure}[t]
    \centering
    \includegraphics[width=0.48\textwidth]{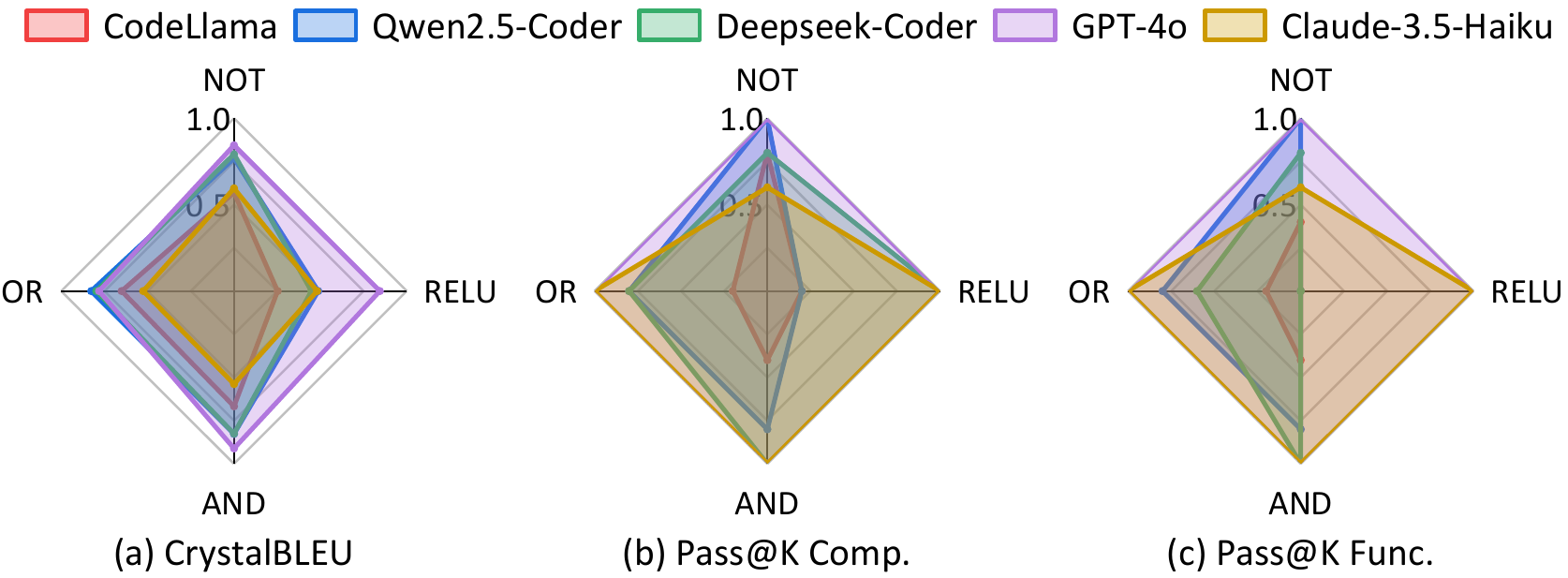} 
    \caption{Comparison of all models across four tasks using Fewshot prompting, using: (a) CrystalBLEU, (b) Pass@k (comp), and (c) Pass@k (func). It significantly improves structural alignment (CrystalBLEU) and correctness for most models, particularly GPT-4o and Claude-3.5-Haiku, while lower-capability models show moderate improvements but still struggle with functional correctness.}
    \label{fig:fewshot-plot}
\end{figure}

\noindent\textbf{RQ3:}
Integrating RAG led to mixed effects on model performance, with minor improvements in some cases but no significant increase in functional correctness. As shown in Fig~\ref{fig:rag-plot}, CrystalBLEU scores increased slightly for certain models, such as Deepseek-Coder on NOT and Qwen2.5-Coder on OR, but performance declined for others, notably Claude-3.5-Haiku on NOT. More importantly, Pass@k (comp) and Pass@k (func) scores remain at or near zero for most models, indicating that access to documentation alone does not significantly enhance compilability or correctness. According to Fig. \ref{fig:other-metrics-plot}, the Wrong Format rate remains high, and repetition error persist, suggesting that RAG does not adequately prevent structural or logical errors.

A possible reason for RAG’s limited success may be that retrieval does not always yield relevant information, as the embedding model is not fine-tuned for TFHE-specific code. This can lead to irrelevant or misleading context being retrieved, which models then misinterpret or misuse, causing hallucinations or incorrect function calls. Consequently, while documentation access is helpful in some cases, it does not provide the structured guidance necessary to ensure functionally correct code.

Across all models and tasks, GPT-4o continues to exhibit the highest performance, maintaining 1.0 Pass@k (comp) and Pass@k (func) for all tasks. However, even with RAG, smaller models like CodeLlama and Deepseek-Coder struggle to produce compilable outputs, frequently misusing API calls or generating incorrect syntax. ReLU remains the most challenging function, with low CrystalBLEU scores across all models and consistently low Pass@k scores.

As illustrated in Fig~\ref{fig:fewshot-plot}, few-shot prompting demonstrates a more substantial impact on code correctness compared to RAG, leading to significant increases in all metrics across most of the models. 

The improvement is especially pronounced in cases where models previously failed completely under baseline conditions. Deepseek-Coder on AND improves from 0.12 to 0.83 (CrystalBLEU) and from 0.0 to 1.0 (Pass@k), demonstrating that few-shot examples significantly enhance structural fidelity and correctness. However, lower-capability models such as CodeLlama still struggle with repetition error, suggesting that while example-based guidance helps, some models have intrinsic limitations in iterative refinement.

Despite these improvements, ReLU remains the most difficult task, as seen in relatively lower CrystalBLEU scores across all models except GPT-4o. The lack of well-established TFHE implementations for ReLU could be a contributing factor, making it harder for models to learn effective patterns even when examples are provided.

\begin{figure}[h!]
    \centering
    \includegraphics[width=0.48\textwidth]{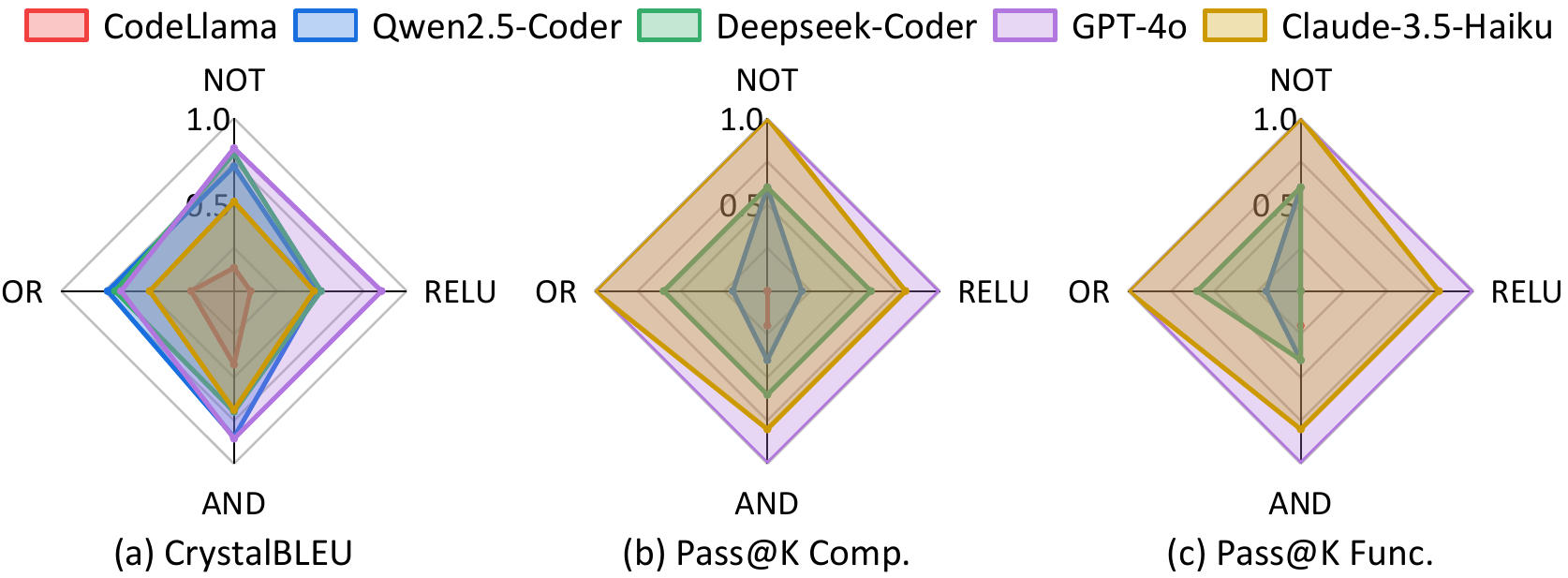} 
    \caption{Comparison of all models across four tasks using the combined RAG + Few-shot prompting technique, using: (a) CrystalBLEU, (b) Pass@k (comp), and (c) Pass@k (func). The combination of RAG and few-shot prompting yields the strongest overall improvements, with GPT-4o and Claude-3.5-Haiku achieving near-perfect correctness across tasks.}
    \label{fig:fewshot-rag-plot}
\end{figure}
Combining RAG and few-shot prompting yields the strongest overall improvements, leading to higher CrystalBLEU scores, reduced error rates, and near-perfect functional correctness for high-performing models. GPT-4o achieves 0.83 CrystalBLEU on NOT, 0.85 on AND, and 0.65 on OR, with 1.0 Pass@k (comp) and Pass@k (func) in most cases, indicating that this combined approach is the most effective.

This combination mitigates errors by ensuring that models receive both correct API references (via RAG) and structured implementation guidance (via few-shot prompting). Notably, Wrong Format and Repetition Error rates are the lowest in this setting, confirming that retrieval works best when paired with explicit examples rather than being used in isolation. 

However, ReLU remains the hardest function across all configurations, with only GPT-4o achieving consistently high performance.

In summary, Few-shot prompting proves more effective than RAG alone, highlighting that explicit examples are more beneficial than documentation retrieval for TFHE code generation. However, the best results occur when RAG and Few-shot prompting are combined, as retrieval helps when supplemented with concrete implementation examples. Higher-capability models like GPT-4o and Claude-3.5-Haiku benefit the most, achieving near-perfect correctness in most tasks, while smaller models continue to struggle with repetition errors and formatting errors despite additional guidance. ReLU remains the most challenging task, with lower success rates across all techniques and models, indicating the complexity of translating ReLU operations into TFHE-based implementations.

\section{Discussion and Futurework}
The results highlight a significant performance gap between proprietary and open-source models in TFHE code generation. GPT-4o consistently outperforms other models, achieving near-perfect correctness across tasks, while open-source models struggle with formatting inconsistencies, repetition errors, and incorrect API usage. Among the tasks, logic gates (NOT, AND, OR) are generally easier to generate, whereas ReLU remains the most challenging due to its arithmetic complexity.

Few-shot prompting proves more effective than RAG alone, significantly improving CrystalBLEU and functional correctness by providing explicit examples. Despite these gains, open-source models continue to struggle with compilability and correctness, suggesting that additional fine-tuning or domain-specific training may be necessary.

\section{Conclusion}
This work presents a novel compiler-integrated framework for generating functionally correct TFHE code from natural language specifications using large language models (LLMs). Our findings highlight several key challenges in LLM-driven cryptographic code synthesis, including persistent formatting errors and repetitions. Our evaluation across multiple LLMs demonstrates that while models such as GPT-4o can reliably generate correct and compilable TFHE code, weaker models require iterative refinement and additional prompting techniques to improve performance. The introduction of retrieval-augmented generation (RAG) and few-shot prompting enhances correctness for smaller models such as Deepseek-Coder, and Qwen2.5-Coder, while CodeLlama struggles to adapt despite multiple refinement cycles, further improvements are necessary to ensure robustness across different models and cryptographic constraints. The results also suggest that LLM-generated TFHE code exhibits varying degrees of structural similarity to reference implementations, with stronger models demonstrating higher fidelity as measured by CrystalBLEU scores. 




\normalem
\bibliographystyle{IEEEtran}
\bibliography{citation}
\newpage
\appendix

\noindent\textbf{Results for Token consumption of models:}

An important aspect of LLM performance in TFHE code generation is token consumption. The baseline results show significant variation in the number of input and output tokens consumed by different models. GPT-4o processed the highest number of input tokens, i.e., 65,829, but generated only 1,784 output tokens, whereas CodeLlama processed 43,639 input tokens and produced 6,838 output tokens. This suggests that larger models like GPT-4o are more selective in their responses, generating more concise outputs, while smaller models tend to produce longer responses, potentially leading to verbosity and errors. Deepseek-Coder had the lowest input token consumption but also the lowest performance, likely due to limited context understanding. The variance in token consumption suggests that token efficiency may be an important factor influencing correctness and structural similarity in TFHE code generation.

When RAG was introduced, input token usage decreased for most models, suggesting that the  retrieved documentation helped constrain response generation. However, this did not consistently lead to improved performance, as irrelevant or incorrect retrievals still contributed to errors. Deepseek-Coder, for example, increased its input token usage dramatically, i.e., from 10,003 to 145,835, but still failed to produce functional TFHE code, suggesting that excessive retrieval may introduce noise rather than useful context. On the other hand, few-shot prompting caused a significant increase in input tokens across all models, as additional example code was included in the prompt. CodeLlama and GPT-4o both show contrasting patterns, where former required more tokens to improve, whereas GPT-4o already generated efficient responses and required fewer tokens. The combined RAG + Few-shot approach further increased token consumption across all models. CodeLlama and Qwen2.5-Coder exhibited the highest input token usage in this setting, but their performance improvements were limited compared to models like GPT-4o and Claude-3.5, which consumed fewer tokens while maintaining high correctness. This suggests that the effectiveness of additional context is model-dependent, with more capable models leveraging fewer tokens to achieve correctness, whereas lower-capability models require more tokens but still struggle with compilability and functional correctness.

\end{document}